\documentclass[conference]{IEEEtran}
\makeatletter
\def\ps@headings{%
\def\@oddhead{\mbox{}\scriptsize\rightmark \hfil \thepage}%
\def\@evenhead{\scriptsize\thepage \hfil \leftmark\mbox{}}%
\def\@oddfoot{}%
\def\@evenfoot{}}
\makeatother
\ifCLASSINFOpdf
\else
\fi
\usepackage[tight,footnotesize]{subfigure}

\usepackage{epsfig}
\usepackage{amsmath}
\usepackage{algorithm}
\usepackage{algorithmic}

 \usepackage[update,prepend]{epstopdf}

\usepackage[square, comma, sort&compress, numbers]{natbib}

\begin{document}
%
\title{Deadline is not Enough: How to Achieve Importance-aware Server-centric Data Centers via a Cross Layer Approach}

\author{
\IEEEauthorblockN{Ming-Hung~Chen, Shi-Chen~Wang, and~Cheng-Fu~Chou\\}
\IEEEauthorblockA{Department of Computer Science and Information Engineering \\
National Taiwan University \\
Email: \{mhchen, wangshichen, ccf\}@cmlab.csie.ntu.edu.tw}}
\maketitle

\begin{abstract}
Today's datacenters face important challenges for providing low-latency high-quality interactive services to meet user's expectation.
For improving the application throughput, recent research works have embedded application deadline information into design of network flow schedule to meet the latency requirement. Here, arises a critical question: does application-level throughput mean providing better quality service?
We note that there are usually a set of semantic related responses (or flows) for answering a query; and, some responses are highly correlative with the query while others do not.
Thus, this observation motivates us to associate the importance of the contents with the application flows (or responses) in order to enhance the service quality.

We first model the application importance maximization problem in a generic network and in a server-centric network. Since both of them are too complicated to be deployed in the real world, we propose the importance-aware delivery protocol, which is a distributed event-driven rate-based delivery control protocol, for server-centric datacenter networks. The proposed protocol is able to make use of the multiple disjoin paths of server-centric network, and jointly consider flow importance, flow size, and deadline to maximize the goodput of most-related semantic data of a query. Through real-data-based or synthetic simulations, the results show that our proposed protocol significantly outperforms D$^{3}$ and MPTCP in terms of the precision at K and the sum of application-level importance.
\end{abstract}


%
\IEEEpeerreviewmaketitle

\section{Introduction}
Over the past few years, deadline-aware services such as web search, online social network, advertising/recommendation system, data warehouse, and online retail systems have been rapidly emerging. Datacenters, as critical computing platforms for ever-growing, high-revenue, must fulfill the requirements of deadline-aware services. Those requirements include: (1) most of deadline-aware services require responsiveness in the sub-second time scale at high request rates, because the shorter and in-time application latency is the key for satisfying user requirement. This is why soft real-time constraints are used for deadline-aware services. (2) Many of those services employ the partition-aggregate operations to parallelly process the massive data sets, which are distributed on thousands of servers, in a divide-and-conquer manner\cite{websearchforaplanet}. Those operations might generate all-to-all and all-to-one communication patterns (e.g., MapReduce\cite{mapreduce}) at the same time. Hence, the datacenter network must be robust enough to cope with those rapid traffic patterns.

Recent works on datacenter networks\cite{bcube, mdcube, mptcp, dctcp} has shown that the key issues of current datacenters are the tree-based infrastructure and the traditional TCP-like transport protocol. The first issue is that the tree-based infrastructure cannot meet the requirements of high-performance large-scale datacenters, which may consist of thousands to millions of servers\cite{mdcube}; and these requirements include: (1) high interconnection bandwidth, i.e., provide multiple disjoin path between servers, (2) low-cost interconnection structure, i.e., only use commodity switches without changes, and (3) low cabling complexity. We note, recently, some server-centric topology designs, e.g., BCube\cite{bcube}, and MDCube\cite{mdcube}, have been proposed to meet those requirements.


\begin{figure}
  \centering
  \subfigure[Conventional TCP]{
  \includegraphics[width=31mm,trim=0 0 0 0, clip]{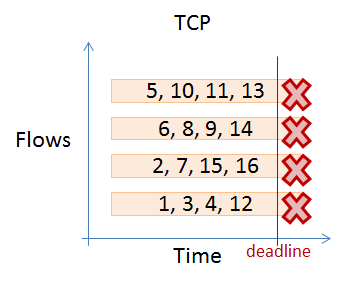}
  }
  \subfigure[Deadline-aware]{
  \includegraphics[width=24mm,trim=0 0 0 0, clip]{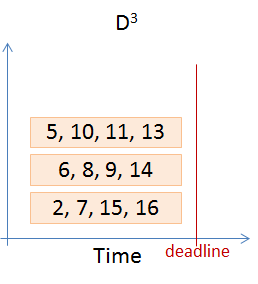}
  }
  \subfigure[Importance-aware]{
  \includegraphics[width=24mm,trim=0 0 0 0, clip]{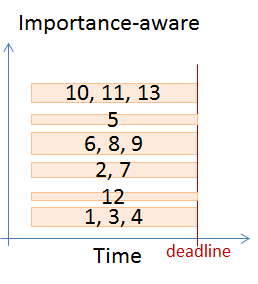}
  }
  \caption[what]{Examples of different delivery protocols}\label{fig:ex3p}
\end{figure}

As for the second issue, the TCP shares the network resources in a laissez-faire manner, which leads to many problems such as TCP Incast\cite{incast2}, low bandwidth utilization when multiple paths exist\cite{mptcp}, and missing deadline\cite{deadline} problems. Many literatures have been studied for coping with the problems of traditional TCP. For example, DCTCP\cite{dctcp} and ICTCP\cite{ictcp} are proposed to mitigate TCP Incast problems. MPTCP\cite{mptcp} is proposed to well utilize multipath infrastructure. HULL\cite{hull} trades a little bandwidth to maintain low network latency for soft real-time services. D$^{3}$\cite{deadline} directly take flow deadline into account. D$^{3}$ proactively suspend some flows to guarantee the remaining flows can meet deadline so as to provide better goodput, i.e., application-level throughput where the flows successfully meet their deadline. With those newly proposed infrastructures and protocol designs, the aforementioned two requirements of deadline-aware services are investigated and fulfilled the robustness of the data center network and the soft real-time constraints.

However, we notice that the robustness of the network infrastructure and the soft real-time constraints of responses cannot totally fulfill the requirements of deadline-aware services. According to recently studies\cite{lagun2011viewser, buscher2010good, cutrell2007you, lorigo2008eye}, the users most concern about is the first few results of deadline-aware services. That is, for the deadline-aware services, the quality of responses, i.e. the precision of top-K results, may significantly affect the degree of user satisfaction on usage experience, and hence this might have great impact on the revenue of the business. This also means that the high goodput, which can be provided by existing deadline-aware protocols\cite{deadline}, is not able to guarantee to get high-quality results for users. For example, in figure \ref{fig:ex3p}, four flows are traveling through a bottleneck switch at the same time, and the number in the flows indicates the rank of response units within the flow. Figure \ref{fig:ex3p}(a) shows that the bottleneck bandwidth cannot let all flows pass through and meet their deadlines. Figure \ref{fig:ex3p}(b) illustrates the behavior of D$^{3}$, which allows requests to be passed through the data center in a first-come first-serve (FCFS) manner. Since the bottleneck bandwidth could support only three flows, one of the flows will be dropped to allow other flows to meet deadline. If the flow with responses of rank 1, 3, and 4 is dropped, the results will has poor quality for a user. Figure \ref{fig:ex3p}(c) demonstrates an example of our heuristic protocol, only several low-rank response units, i.e., rank 14, 15, and 16, are dropped and all high-rank response units are delivered before the deadline. That is, although the network bottleneck exists, the deadline-aware services can still provide high quality results to users. Hence, we believe how to provide high-rank results for users is an important issue toward high-quality deadline-aware services; and, our ideas for design of an importance-aware datacenter network are as follows:

\begin{enumerate}
\item Application-level importance is associated with each response unit, not flows, while deadline is associated with flows. The flows (and the response units within it) are valid only when they arrive before the deadline.
\item The flow size may vary from less than 10KB to more than 500KB because of the different applications. That is, the protocol must be aware of the size of the flows as well.
\item Some of the flows are short and so is their deadline. In addition, the RTT is small in a datacenter. So, the reaction time of the delivery control protocol must be short enough to cope with those short and near-deadline flows. That is, centralized or heavy-weight mechanisms are impractical.
\item The application-level importance of all applications in a datacenter must be normalized to provide a fair-sharing network. However, this can be measured and normalized manually or automatically by datacenter administrators.
\end{enumerate}

In this work, we exploit the application-level content information, i.e., data importance, to meet the most important requirement, i.e., the precision of top-K results, which affects user experience on deadline-aware services, To figure out the optimal solution and performance upper bound, we first model the datacenter network as a flow network for analyzing the application importance maximization problem with the constraints of network link capacity, flow conservation, and flow deadline.
However, due to the complexity of the optimal solution, we do not consider it can be put into practice.

Therefore, we further propose a novel distributed rate-based importance-aware delivery control protocol. The basic idea of our protocol is to greedily choose a set of the flows in the networks that can contribute the most importance before deadline under the constraint of the network capacity. The information of flows travelling on a switch, i.e., the deadline, remain size, averaged data importance in the flow, sending rate of each flow, is known by all servers connected to the switch at the initiation stage of the flows. Hence, each server can estimate if its upcoming flow will contribute more data importance than existing flows by a new proposed metric jointly considering remain size, remain transmission time, and averaged data importance of a flow, such that we can suspend less-important flows and replace them with new upcoming flow. 
To utilize the multiple disjoin paths while maximize application data importance, we choose the combination of paths and rates that affects less data importance in all available disjoin paths to the destination. Notice that the deadline of flows and traffic amount in a datacenter vary over time, and this means some of the suspended flows may be recovery later for them, which can still meet their deadlines. Hence, once a flow has completed, the source of the flow will inform the owners of the suspended flows to see if any of them can recover flows before their deadline. To further improve the delivery ratio of highly important response units, we use statistical clustering algorithm, i.e., k-mean algorithm, to cluster responses of a flow into two (or more) small flows based on the importance of responses. That is, most of highly important response units of any requests will be delivery in deadline while less-important responses will be ignored for improving user perceived service quality.
In summary, the advantages of the proposed protocol include:

\begin{enumerate}
\item We design the proposed protocol in a fully distributed manner, i.e., no centralized controller is required.
\item The proposed protocol adaptively adopts the most beneficial flows on the fly and recovers less-important flows when network resources are available.
\item We split the original flow into two (or more) small flows by statistical clustering algorithm, i.e. the k-means algorithm, for further improving the successful delivery ratio of important response units.
\item The proposed protocol is able to make use of multiple disjoin links and balances the load between links for enhancing the throughput.
\item No expensive or customized hardware is required, i.e., only commodity switches without changes are used.
\end{enumerate}

The results based on real data show that the proposed algorithm can provide better application-level user perceived performance than MPTCP\cite{mptcp} and D$^{3}$\cite{deadline}, i.e. the precision at K. That is, we believe our protocol can better fulfill the requirements of deadline-aware services for providing better user experience, and hence producing more revenue for the providers.

%
%
%
%
%
%
%
\section{Problem Formulation}
In this section, we formulate the application importance maximization problem for the optimal solution, which also means to find the maximal sum of the importance of flows in the network which meet deadline. We first model the importance-aware datacenter networks and formulate the problem. The complexity of the optimal solution in server-centric networks is then addressed. In summary, we found that the problem cannot be solved in polynomial time. Therefore, we present a distributed heuristic protocol based on similar ideas of the optimal solution.

\subsection{Model of Importance-aware Data Center Networks}
We consider a data center network is represented by G=(V,E), where $V$ represents a set of nodes, i.e., servers or switches, and $E$ represents the links between nodes. The bandwidth capacity of each link $(u,v)\in E$ is $C(u,v)$. There are $N$ flows, i.e., $F_{1}, F_{2},...,F_{N}$, in the data center network, and each flow has its source $s_{i}$, destination $t_{i}$, begin time $b_{i}$, deadline $d_{i}$, and a set of response units $R_{i}$. For each response $r$ in $R_{i}$ of flow $F_{i}$, the importance of $r$, i.e., the similarity to the query of $R_{i}$, is denoted as $m_{i}$, and the size of each response $r$ is denoted as $rs_{r}$. Therefore, flow size $Fs_{i}$ is equal to the sum of the response unit size of $R_{i}$. We then denote the averaged importance of $F_{i}$ as $I_{i}$=$\frac{\sum_{r\in R_{i}}m_{r}}{\|R_{i}\|}$. Consequently, we regard the flow in the importance-aware network as a five-tuple vector, i.e., $F_{i}$=($s_{i}$, $t_{i}$, $d_{i}$, $Fs_{i}$, $I_{i}$). The transmission rate of $F_{i}$ on link $(u,v)$ at time $t$ is represented by $r_{i}^{T_{j}}(u,v)$.

\subsection{Global Optimization in Data Center Networks}

In a data center network, once a flow cannot meet its deadline, all response units of that flow are regarded as lost. This property makes how to maximize application importance becomes difficult. Hence, we split each flow into several tiny flows, and each tiny flow contains a single response unit. That is, the new flow set $N'$ consists of $\sum_{i=1}^{N} \|R_{i}\|$ flows. Since each new flow only contains one smallest unit that contribute importance, we do not allow them to be split anymore to guarantee they can be completely delivered. The new flows can also be defined in the form of five tuples as: $F_{i}$=($s_{i}$, $t_{i}$, $d_{i}$, $Rs_{i}$, $m_{i}$). Assume the end time of the flow $F_{i}$ is $e_{i}$, the first begin time of the flows is $T_{p}$, the last end time of the flows is $T_{q}$, and the transmission and queueing delay of flow $F_{i}$ is $\bigtriangleup_{i}^{delay}$, we can define the application importance maximization problem as:
\begin{equation}
\max \sum_{k \in M} m_{k} ,
\label{eq:objective}
\end{equation}
\begin{equation}
\begin{split}
M = \{ i | 1 \leq i \leq N', \frac{ \sum_{j=T_{p}}^{T_{q}-1} (( \sum_{u \in V} r_{i}^{T_{j}} (s_{i}, u) ) \times (T_{j+1} - T_{j} ))} { e_{i} - b_{i} } \\
\geq \frac{ds_{i}}{d_{i} - b_{i} - \bigtriangleup_{i}^{delay} } \}
\label{eq:Mset}
\end{split}
\end{equation}
subject to:

\begin{enumerate}
\item Link capacity constraint: for each link $(u,v)$ at time $t$, the total flow rate on the link must be equal to or smaller than the link capacity.
\begin{equation}
\sum_{i=1}^{N'} r_{i}^{t} (u,v) \leq C(u,v)
\label{eq:capacity}
\end{equation}
\item Flow conservation constraint: the amount of in-flow and out-flow must be the same on all nodes unless the source or the destination of the flow is the node.
\begin{equation}
\sum_{w \in V} r_{i}^{t} (u,w) = 0 , u \neq s_{i}, t_{i}
\label{eq:conservation}
\end{equation}
\end{enumerate}

We notice that the application importance maximization problem can be reduced to an unsplittable flow problem. In the unsplittable flow problem, a $n$-vertex graph $G(V,E)$ with edge capacity $c_{e}$ and the set of $k$ vertex pairs $T={(s_{i},t_{i}),i=1,...,k}$ are given. Each pair ($s_{i},t_{i}$) in $T$ has a demand $\rho_{i}$ and a weight $w_{i}$. The goal attempts to find the subset of pairs from $T$ with the maximum weight so that entire demand for each of such pair can be routed on its path under the link capacity constraint. Now we suppose all flows in the network has the same begin time and deadline, which represents a simplified case of the maximization problem. We can map the set of $N'$ flows to the set of $k$ vertex pairs, the minimal averaged transmission rate $r_{i}^{d_{i}-b{i}}(s_{i},t_{i})$ that meets its deadline to the demand $\rho_{i}$, and the importance of the flow $m_{i}$ to the weight $w_{i}$, so as to convert this application importance maximization problem to the unsplittable flow problem. That is, the result set of the unsplittable flow problem denotes the flows that maximize the application importance. This also represents that for the flows whose demand cannot be satisfied in the unsplittable flow problem, they should be dropped in advance because they will degrade the sum of the demand, i.e., the sum of application importance. However, the unsplittable flow problem is regarded as NP-complete\cite{approximationunsplittable, reducibility}, and solving the unsplittable flow problem needs global information, i.e., a centralized system is required. Those limitations make this solution become impractical for datacenter networks.

\subsection{Local Optimization in Server-centric Datacenter Networks}

One of the critical problem in the unsplittable flow problem is to find the path for flow $F_{i}$ that meets its demand, i.e., the minimal averaged transmission rate. It is difficult to solve because in a datacenter network, many intersecting path may cross with each other on nodes in a network. Finding those link-joint path makes the allocation of capacity to each flow become complicated. We note that the server-centric network architectures not only have many advantages over traditional network architectures; and, one of their properties is to provide fixed number of link-disjoin paths between servers. This property can help us on reducing the complexity of the application importance maximization problem, and we will explain the details as follows.

Our basic idea is to allow each server to determine whether it can start a new flow by traversing through the disjoin paths to the destination. If (a) the smallest remain capacity of the paths is insufficient to meet deadline, and (b) the importance of the new flow is less than any other flows on the paths, the flow will be suspended till there are sufficient available capacity to meet its deadline. If the new flow can contribute more importance than existing flows, it will notify the less-important flows to be suspended and starts transmission. That is, each server can decides if a flow should be transmitted according to the partial information of the network.

After the disjoin paths are given, we notice that the applications importance maximization problem can be reduced to a multiple knapsack problem. We first map the available link capacity $C(u,v)$ of link $(u,v)$ to the capacity of knapsack $W_{(u,v)}$. The flow $F_{i}$ is mapped to the item $x_{i}$ that are going to be place in the knapsack. The importance of the flow $F_{i}$, $m_{i}$, is mapped to the value of the item as $v_{i}$. The minimal averaged transmission rate $r_{i}^{d_{i}-b{i}}(s_{i},t_{i})$ that meets its deadline is mapped to the weight of the item $w_{i}$. We then define the set of disjoin paths between $s_{i}$ and $t_{i}$ that flow $F_{i}$ will pass through as $L_{i}$. This set of links $L_{i}$ can be treated as a set of knapsack sets $K_{i}$ of flow $F_{i}$. A knapsack set of $K_{i}$ consists of knapsack $k{(s_{i},m)},...,k{(n,t_{i})}$, which represents the available capacity of a disjoin hop-by-hop path from $s_{i}$ to $t_{i}$ and all edges on this path are denoted as $Ke$ for latter use. We use $k_{i,(m,n)}$ to denote this knapsack set and use smallest available capacity of the knapsack in the set as the capacity $W_{i,(m,n)} = \min W_{(u,v)}, (u,v)\in Ke$. Once a item $x_{i}$ is placed in a knapsack set $k_{i,(m,n)}$, we denote it as $x_{i,(m,n)}$, and at the same time the available capacity of all knapsacks in the set will be reduced by $w_{i}$. Therefore, assume there are $N$ flows sent from the source node, we can reduce the local application importance maximization problem as follows:

\begin{equation}
\max \sum_{i=1}^{N} v_{i}x_{i}
\label{eq:knapsack}
\end{equation}
subject  to
\begin{equation}
 \sum_{i=1}^{N} w_{i}x_{i,(m,n)} \leq W_{i,(m,n)}
 \label{eq:weightContraint}
\end{equation}
\begin{equation}
\begin{split}
x_{i} = \left\{\begin{matrix} 1, if x_{i,(m,n)} = 1, \exists (m,n) \in L_{i} \\
0, otherwise
\end{matrix}\right.
\end{split}
\label{eq:xi}
\end{equation}
\begin{equation}
x_{i,(m,n)} \in \left \{ 0,1 \right \} \forall (m,n) \in  E
\label{eq:x1}
\end{equation}

The multiple knapsack problem, however, is known as a NP-complete problem\cite{reducibility} so it requires very long computational time while the flows are used to be short and massive. Even if the information of all upcoming flows are known, the problem can be solved in pseudo-polynomial time; but, this is still impractical in datacenter networks.

\section{Importance-aware Delivery Framework for
Server-centric Networks}

In this section, we describe our heuristic importance-aware delivery protocol for server-centric datacenter networks. The design goals of our protocol include:
\begin{enumerate}
\item Maximize application importance, i.e., providing high quality results for deadline-aware services by finishing most important flows before deadline.
\item Light-weight and fully distributed, i.e., fast enough to cope with datacenter networks and no need of customized or expensive hardware.
\item High network utilization, i.e., exploiting all available static multi-disjoin paths of server-centric networks.
\end{enumerate}
The key insight guiding our protocol design is: given (a) the topology information, (b) the size and importance of each response unit in a flow, as well as (c) the deadline of a flow, each server in the server-centric networks can determine (1) which part of the flow is important, (2) the rate to meet the deadline constraint. The servers, then, can determine if the rate should be allocated according to the importance of the flow and the network status of the paths to the destination.

\subsection{Flow Importance Contribution Metric}

Before introducing our protocol, we must address a fundamental issue: How to determine whether a flow is important? If all flows begin at the same time and they have the same deadline and size, the answer is quite simple. We use $I_{i}$ of flow $F_{i} $to decide if it can contribute more importance than other flows. However, in the datacenter network, the flows come or leave all the time, and their deadline, size, and importance are different. Thus, we need a new metric to determine the importance of a flow.
Our idea is to measure the carried importance per data unit per time unit. This is because after some transmission time, the remaining size of a flow will become small. However, only when the remaining part of the flow is fully delivered, the importance of the flow can be count in. That is, the metric must take the remaining size and remaining time of the flow into account in by considering temporal and spatial diversity together. Hence, we imposes the averaged flow importance $I_{i}$, remaining size $RS_{i}$, and remaining transmission time $RT_{i}$ to compute the flow importance contribution (FIC) of flow $F_{i}$. The FIC of flow $F_{i}$ is defined as:

\begin{equation}
 FIC_{i}=\frac{I_{i}}{RS_{i} \cdot RT_{i}}
 \label{eq:ficofi}
\end{equation}

\subsection{Protocol Overview}

We first give a overview of the protocol. Our protocol is a distributed event-driven rate-based delivery control protocol, which means we assume the datacenter network is able to assign each flow by a fixed transmission rate on multiple paths till (1) a server tries to initial a new flow, (2) a flow ends, hence some capacity becomes available and each individual server will determine if: (a) Some of its suspended flows can be awaken, for they are more important and the available capacity are sufficient to meet their deadline, and (b) some of its flows can increase their transmission rate to shorten their transmission time


To improve the delivered application importance, we further split a flow into two (or more) small flows by the statistical clustering algorithm, e.g. the k-means algorithm, for improving the successful delivery ratio of more important response units. The details of the protocol are in the following.

\subsubsection{Distributed Rate Control}

\begin{figure}
  \centering
  \includegraphics[width=80mm]{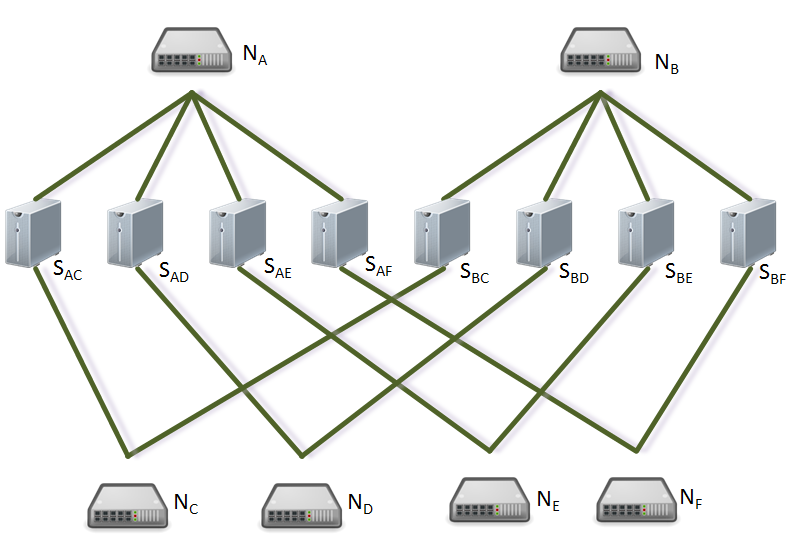}\\
  \caption{A simplified 2-level BCube topology}\label{fig:bcube}
\end{figure}

In the server-centric datacenter networks, several servers are connected by a commodity network component, i.e., a switch, as a set of servers. Those servers may connected to other servers via another network component to form a larger network. Hence, we define the neighbors of a server as the other servers connecting with the server by the same switch. For example, figure\ref{fig:bcube} shows a simple BCube\cite{bcube} network, in which server $S_{AC}, S_{AD}, S_{AE}, S_{AF}$ are connected by the same switch $N_{A}$, and server $S_{AC}, S_{BC}$ are connected by $N_{C}$. Therefore, the neighbors of $S_{AC}$ include $S_{AD}, S_{AE}, S_{AF},$ and $S_{BC}$. With our protocol, each server maintains the remaining link capacity of all links to its neighbors via connected network components. It also maintains the information of the flows that traverse through its neighbors or itself, which includes the deadline of flow, the transmission rate of flow, and the averaged importance of the flow.
With those information, the servers connected to the same switch can carefully allocate the rates of the links of the switch, so as to prevents oversubscribing the capacity of each link, which could result in building up packet queue on the switch, packet loss, and making flows missing deadline.

\subsubsection{Flow Initialization}

As the deadline-aware applications expose flow size, deadline, and importance information when initializing a flow $F_{i}$, the server can calculate the flow's importance contribution $FIC_{i}$ and minimal transmission rate $r_{i}^{d_{i}-b{i}}(s_{i},t_{i})\simeq\frac{Fs_{i}}{d_{i}-b{i}}$. Then, the server looks up the shortest equal-cost disjoin paths (and the second shortest paths if only one shortest path exists) to the destination for the next hops. The rate request is estimated at the server based on the flow information of its neighbors. If the network is under-loaded, the rates of existing flows with lower $FIC$ than $FIC_{i}$ will be allocated till the rates can satisfy the minimal transmission rate of the flow. If the network is over-loaded, the existing flows will be asked to release the extra rates they occupied other than their minimal rates. If the allocated rate is still insufficient to satisfy the flow, the network will be treated as over-loaded and remaining rate will be allocated from existing flows with lower $FIC$ than $FIC_{i}$. However, sometimes the network is light-loaded. In this case, after all flow successfully allocates its minimal transmission rate, the remaining capacity will be assigned to each flow proportion to its $FIC$.

Once the initial rate request is succussed at the server, the request will be updated and forwarded to the next hops, i.e., the relay nodes, to further allocate the rate from the next hop to the next two hops, and so on. The demanded rate and available $FIC_{i}$ of a path, which is recorded in the rate request, is proportion to the estimated available capacity of the paths based on the source's local information to balance the load between paths. To maximize the application importance by trading less-important flows with most-important flows, we only allow the new flow that its FIC is less than the sum of FICs of the affected flows on each node to be initialized. That is, each server along the path will allocate the most rate given available $FIC$. The rate request will be processed and forwarded to the next hop till reaching the node on the same switch of the destination. Notice that on each node, the allocation should be done twice, i.e., to allocate the rate from the node to the network component, and the rate from the network component to the next hop. The detailed rate allocation algorithm can be found in Algorithm \ref{alg:rate-allocation}.

After the server receives the responses of the rate request, it can determine the available capacity of a path for flow $F_{i}$ by the smallest available rate reported by the nodes on the path. Therefore, if the sum of the smallest available rate of all paths is less than the minimal available rate, the flow should not be transmitted. Otherwise, the server will allocate the flow rate proportion to the available capacity of each path, and broadcast the rate allocation notification to all neighbors. The notification will be re-broadcasted by the relay nodes until reaching the destination. Besides, for some existing flows might be suspended, the on-path node on the same switch of the flows will take the responsibility to inform the sources of to-be-suspended flows to stop the transmission. The information of the suspended flows is maintained by this node for further flow recovery. Note that the RTTs in the datacenter network is very small ($\approx300\mu$s), thus the protocol should take very short time to allocate the rate and suspend existing flows.

\begin{tiny}
\algsetup{indent=2em}
\renewcommand{\algorithmicrequire}{\textbf{Input:}}
\renewcommand{\algorithmicensure}{\textbf{Output:}}
\newcommand{\RECORD}{\textbf{Recorded Info:}}
\newcommand{\factorial}{\ensuremath{\mbox{\sc Factorial}}}
\begin{algorithm}[h!]
\caption{Rate Allocation on Each Node} \label{alg:rate-allocation}
\begin{algorithmic}[1]
\REQUIRE minimal transmission rate $r_{desired}$, the link set $L_{tr}$ to the nexthop, available $FIC_{f}$ of the flow of the path.
\ENSURE allocated rate for the flow $r_{allocated}$.

\RECORD remaining capacity of the links in $L_{tr}$, flow information in $L_{tr}$.
\medskip
\WHILE {$L_{tr}\neq \phi$}
\STATE $L_{cur} \leftarrow$ GET-NEXT-LINK($L_{tr}$)
\IF {remaining link capacity in $L_{cur} \geq r_{desired}$}
\IF {remaining link capacity - $r_{desired} >$ 0 }
\STATE $r_{add} \leftarrow$  assign spare capacity in proportion to $FICs$
\ELSE
\STATE $r_{add} \leftarrow 0$
\ENDIF
\STATE $rt_{L_{cur}} \leftarrow r_{desired} + r_{add} $
\ELSE
\STATE $F \leftarrow$ {existing flows traversing through $L_{cur}$}
\STATE $R  \leftarrow$ desired rate $r_{desired}$
\STATE $FIC \leftarrow$ available FIC of the path $FIC_{f}$;
\WHILE { $F \neq \phi$}
\STATE $f_{i} \leftarrow$ EXTRACT-MIN-FIC(F)
\STATE $r_{i} \leftarrow$ transmission rate of the flow $f_{i}$
\STATE $fic_{i} \leftarrow$ FIC of the flow $f_{i}$
\IF { $ FIC < fic_{i}$}
\STATE $rt_{L_{cur}} \leftarrow r_{desired} - R$
\STATE \textbf{break};
\ENDIF
\STATE $R \leftarrow R - R_{i} $;
\STATE $FIC \leftarrow FIC - fic_{i}$
\ENDWHILE
\ENDIF
\ENDWHILE
\STATE $r_{allocated} \leftarrow$ MIN($rt_{(*)}$)
\RETURN $r_{allocated}$;
\end{algorithmic}
\end{algorithm}
\end{tiny}

For example, we let the $S_{AD}$ as the flow source and $S_{BE}$ as the flow destination. The minimal transmission rate of the new $F_{i}$ is 120kbps. The shortest disjoin paths between $S_{AD}$ and $S_{BE}$ are $P1=\{S_{AD}, (N_{D}), S_{BD}, (N_{B}), S_{BE}\}$, $P2=\{S_{AD}, (N_{A}), S_{AE}, (N_{E}), S_{BE}\}$. Next, $S_{AD}$ will estimate the available link capacity between ($S_{AD}$, $S_{BD}$) $=$ 180kbps , and ($S_{AD}$, $S_{AE}$) $=$ 220kbps. Since the sum of the available capacity is greater than the minimal transmission rate, the flow can be transmitted to the next hop. $S_{AD}$ then sends the rate requests to the relay nodes $S_{BD}$ and $S_{AE}$ to allocate $\frac{120\ast180}{180+220}=54$kbps on path $P1$ and hence 66kbps on path $P2$. Because nodes $S_{BD}$ and $S_{AE}$ are on the same switches of the destination $S_{BE}$, the rate request no longer needs to be forwarded. Now suppose the smallest available rate of path $P1$ is 60kbps and of the path $P2$ is 90kbps. $S_{AD}$ will allocate $\frac{120\ast60}{60+90}=48$kbps on path $P1$, and hence 72kbps on path $P2$.

\subsubsection{Flow Completion and Flow Recovery}

When a flow finishes its transmission, a notification will be broadcasted and forwarded, which is similar to the final procedure of the flow initialization. The notification is to tell the nodes along the path and all of their neighbors that some network resources are released. Hence, those nodes can determine if (1) some suspended flows can be awaken, and (2) the transmission rate of existing flows can be increased. This also triggers the nodes along the path that suspended flows for the ending flow to inform the sources of the suspended flows for recovering. The flow recovery and rate adjustment procedure is exactly the same as the aforementioned flow initialization procedure.

The main reason we allow the recovery of suspended flows is, in the datacenter network, the deadline of flows is different. Hence, some suspended flows may be recovered because they have longer deadline. That is, flow recovery can further help to maximize application importance.


\begin{figure}
  \centering
  \includegraphics[width=88mm,trim=80 0 80 0, clip]{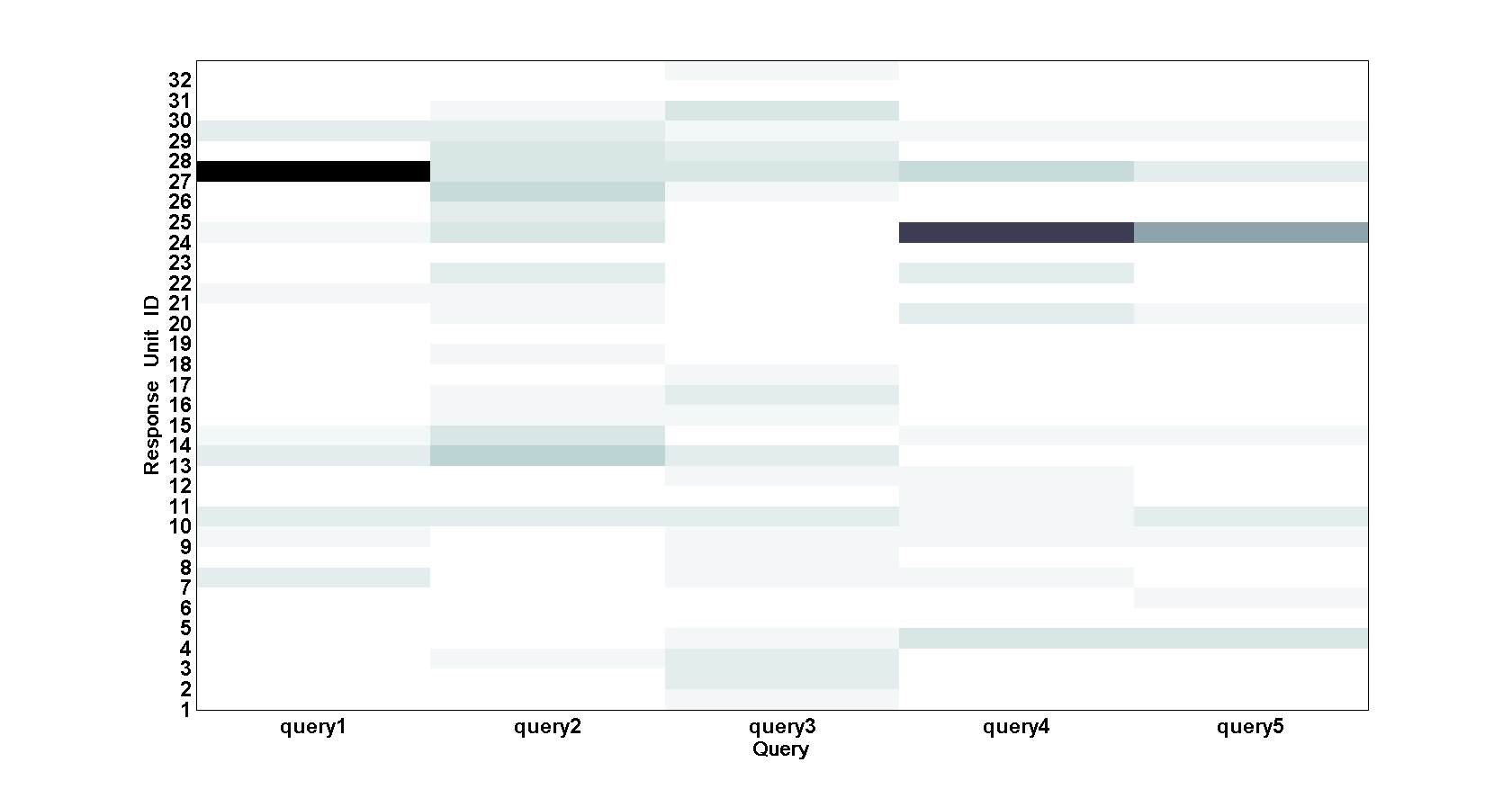}\\
  \caption{Importance distribution of flows from Host \#93}\label{fig:imstat}
\end{figure}

\subsubsection{Flow Splitting}

In Section II, we split each flow into tiny flows that each of them contains only one single response unit for maximize the application importance. However, it is impractical to enforce the same approach in real networks because it might generate massive amount of flows. However, we notice that in the datacenter, the datasets of deadline-aware services are mostly uniform distributed over a lot of servers. Therefore, for a general query, the averaged importance of the response flows could be similar. This makes it hard to distinguish which flow is more important than others. We note that for a general query, a server will generate a few results with high similarity, and others with low similarity. Figure\ref{fig:imstat} shows the importance distribution of the response units within the flows. The response unit with deep color represents high correlation between this response unit and the corresponding query. This result is generated based on the real-world dataset from NII Test Collection for IR Systems (NTCIR) Project\cite{ntcir} and the dataset are distributed to the hosts based on the data placement policy of GFS\cite{gfs}. Therefore, we believe the phenomenon can also be found in many deadline-aware services. Based on this observation, we employ statistical clustering algorithm, i.e., k-mean algorithm, to divide the flow into several small flows. How many parts a flow should be split is ought to be determined by the application administrators and it is beyond the scope of this work. Here, we simply split a flow into two smaller flows, one of them consists of most important response units, and the other consists of remaining response units. Indeed, the most important flow has higher $FIC$ than the other. Thus, those most important small flows will be transmitted prior to other flows. This further increases the successful delivery ratio of most important response units, and hence help to maximize application importance.

\section{Performance Evaluation}

We implement a three-level BCube network architecture\cite{bcube} with parameter $k=5$ as the server-centric datacenter infrastructure on the event-driven network simulator ns-2\cite{ns2}. The proposed protocol, $D^{3}$\cite{deadline}, and MPTCP\cite{mptcp} are implemented as the delivery protocols on BCube topology as well. The MPTCP implementation is based on the open-source project multipath-tcp on ns-2\cite{mptcpns2}. As mentioned in \cite{bcube}, a host within $n$-level BCube connects to $m$ switches and at least $m$ disjoin path between any host pair is guaranteed. Hence, the BCube network consists of 125 hosts, and each host connects to 3 switches. To simulate the common commodity switches used in datacenter networks, we set the packet buffer to 4MB at the switch, and the packet size is 1KB. Each switch can buffer at most 4000 packets before dropping any packet. However, because the proposed protocol is based on the rate control mechanism, the size of the queue are small ($<$200) all the time. The link capacity is 1Gbps. The round trip propagation delay varies from 35$\mu$s to 100$\mu$s depending on the hop counts between the pair of nodes. We follow the deadline setting as in \cite{deadline}, where the mean value of deadline of flows is set to 20 milliseconds, 30 milliseconds, and 40 milliseconds, which represent the emergency of response from tight, moderate to loose. As mentioned in Section III, each flow is divided into two smaller parts: an important flow and a regular flow. The traffic pattern is similar to the traffic generated by partition-aggregate operations. An aggregating host is first randomly selected from the network, and the rest of 124 hosts will then generate flows to the aggregating host at the same time.

We use two datasets to conduct the evaluation. The first one is a synthetic dataset. In this scenario, the distribution of flow size follows the uniform distribution as in \cite{deadline}. The flows in the light-load network is uniformly distributed across [2KB, 50KB]. Flow size across [50KB, 100KB] and across [100KB, 150KB] represent medium-load and heavy-load networks respectively. For the distribution of flow importance, we assume half of the response units of a flow have high importance, which is set to 10, and the rest of response units have low importance, which is set to 1. The other dataset consists of a set of real-world articles, which comes from NII Test Collection for IR Systems (NTCIR) Project\cite{ntcir}. We distribute the data to the hosts according to the data placement policy of GFS\cite{gfs}.

\subsection{Metrics}

The primary goal of our evaluation is to determine the value of employing flow importance information to allocate network capacity. We would like to verify if the proposed protocol is able to exploit path diversity. Thus, the following metrics are used to conduct the evaluation on the synthesis dataset.

\begin{enumerate}
\item Application-level throughput: the sum of meeting-deadline flow size. Only the flow delivered before its deadline can be count in.
\item Application-level aggregated importance: total amount of flow importance delivered before flow deadline. Flows with high importance contribute important responses to users. So, the aggregated application importance represents the quality level users experienced.
\item Ratio of meeting-deadline flows: the ratio of flows delivered before their deadline. This results can verify if highly important flows of the proposed protocol does have better delivery ratio.
\end{enumerate}

For the evaluation on the real dataset, we use the precision at $K$, which is the fraction of received response units that are relevant to the query, to determine if the proposed protocol can provide more highly important responses to the users. The formal definition of precision at $K$ is:

\begin{equation*}
 \textit{Precision@K}=\frac{|\{\textit{Top-K relevant units}\}\bigcap\{\textit{received units}\}|}{|\{\textit{received units}\}|}
 \label{eq:patk}
\end{equation*}

\subsection{Compared Protocols}

We compare the proposed protocol with $D^{3}$\cite{deadline} and MPTCP\cite{mptcp}. MPTCP exploits the path diversity in the datacenter networks to achieve high network utilization. It adopts TCP-like transport protocol and does not consider flow deadline and importance. $D^{3}$ leverages explicit rate control mechanism in a semi-distributed fashion to enforce the deadline of the flows. Although it takes flow deadline into account, $D^{3}$ does not be aware of the path diversity in the datacenter. Therefore, we modify $D^{3}$ to randomly select one of the disjoin paths for a flow to improve the network utilization and balance the load of paths.

For a fair comparison, we improve $D^{3}$ and MPTCP by injecting the flow splitting mechanism as mentioned in Section III.B.4. The flow splitting mechanism helps $D^{3}$ and MPTCP to enhance the delivery probability of highly important response units.

Since the performance of original $D^{3}$ and MPTCP are worse than the modified versions, we only present the results of modified $D^{3}$ and MPTCP here, due to the lack of space.

\begin{figure}
  \centering
  \subfigure[In light-loaded network]{
  \includegraphics[width=95mm,trim=25 0 20 0, clip]{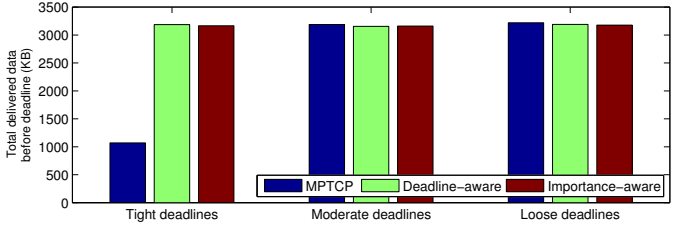}
  }
  \subfigure[In medium-loaded network]{
  \includegraphics[width=95mm,trim=20 0 20 0, clip]{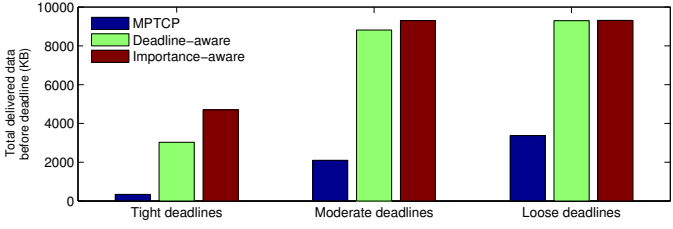}
  }
  \subfigure[In heavy-loaded network]{
  \includegraphics[width=95mm,trim=20 0 20 0, clip]{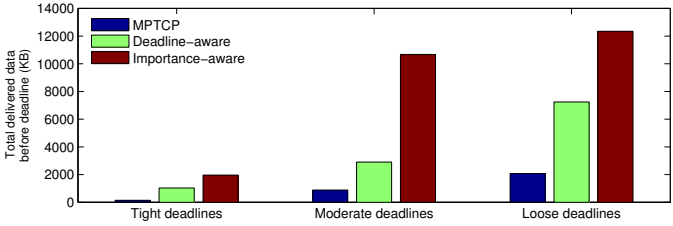}
  }
  \caption[what]{Goodput under different deadlines}\label{fig:gp}
\end{figure}

\subsection{Results of synthesis dataset}

Figure \ref{fig:gp} gives the application-level throughput of the protocols. Figure \ref{fig:gp}(a) shows the goodput in the light-load network. Both the proposed protocol and D$^{3}$ can mostly cope with all flows and both of them outperform MPTCP. MPTCP aggressively increases transmission rate till packet loss occurs. Once a packet is lost, MPTCP needs to wait for a interval of TCP retransmission timeout (RTO) before retransmission. Because the default value of RTO is usually set to 100ms or 200ms, once a packet of a near-deadline flow gets lost, the flow will mostly miss its deadline. This is the main reason why the performance of MPTCP is poor in all cases. D$^{3}$ has similar goodput to the proposed protocol at low (medium) load with loose deadlines. However, as shown in Figure \ref{fig:gp}(b)(c), when the network is under heavy-loaded or when the deadline is tight, the proposed protocol outperforms D$^{3}$. This is because D$^{3}$ does not well exploit path diversity.

As mentioned in \cite{incast2}, at a partition-aggregate operation, massive flows will be generated, and this leads to huge amount of traffic at the links to the aggregating host. For example, when the network is light-loaded with moderate deadlines, the total network capacity to the aggregating host must be equal to or larger than $\frac{(50+2)\div2\ast124\ast8}{0.03\ast1000\ast1000}\simeq0.86$Gbps for a partition-aggregate operation. Since the proposed protocol provides 1.9 times, 3.67 times, and 1.7 times of goodput over D$^{3}$ with tight, moderate, and loose deadlines. We believe the proposed protocol is robust and be able to deal with rapid traffic.

\begin{figure}
  \centering
  \includegraphics[width=95mm,trim=35 0 20 0, clip]{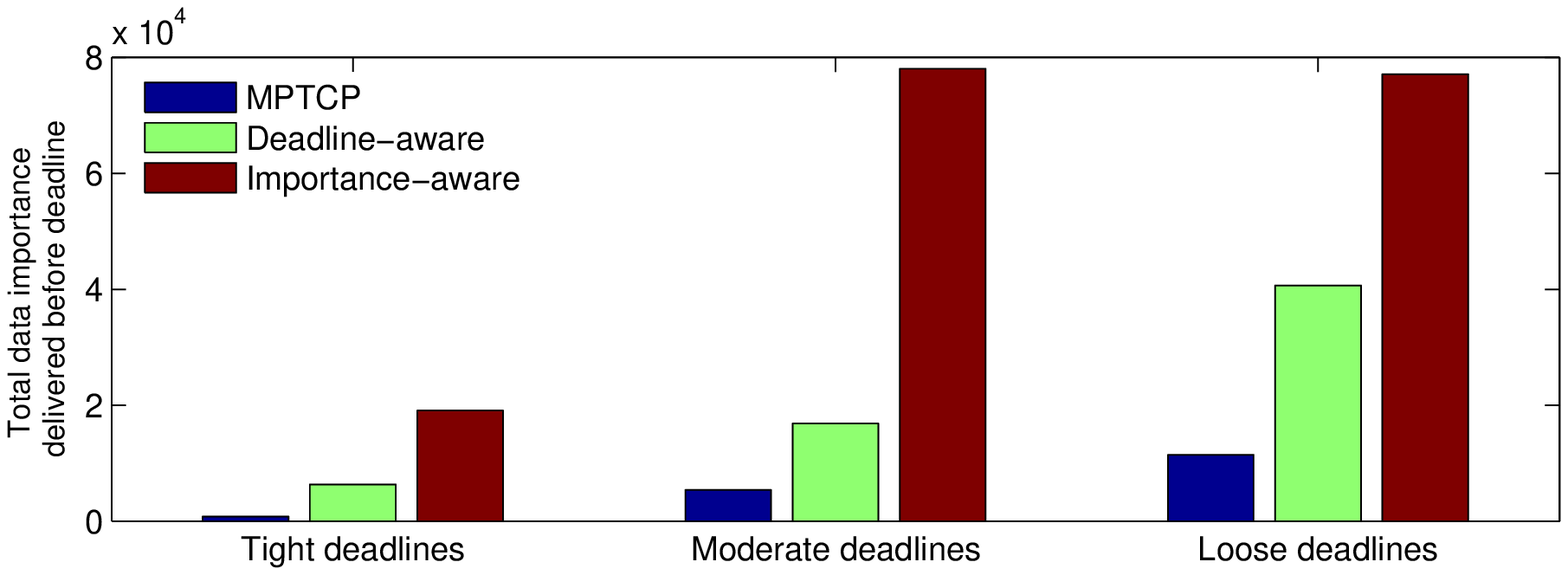}\\
  \caption{Ratio of meeting deadline flows}\label{fig:im}
\end{figure}

\begin{figure}
  \centering
  \includegraphics[width=95mm,trim=35 0 20 0, clip]{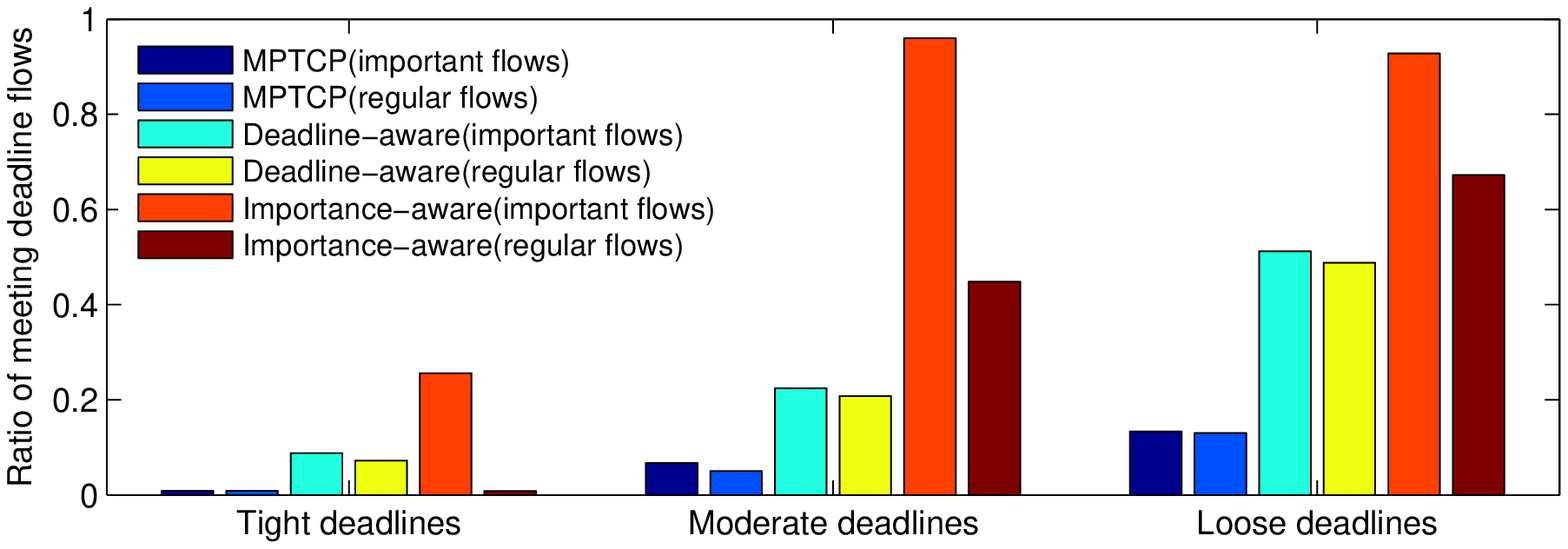}\\
  \caption{Ratio of meeting deadline flows}\label{fig:ratio}
\end{figure}

Figure \ref{fig:im} illustrates the sum of importance contributed by meeting deadline flows. When the network load is heavy, the proposed protocol provide 3.0 times, 4.64 times, and 1.9 times data importance contributions to D$^{3}$ with tight, moderate, and loose deadlines. Note that the difference in total importance is larger than in goodput. That is, when the network capacity becomes the critical bottleneck, the proposed protocol can well utilize available bandwidth to deliver highly important flows. Hence, the total data importance will be greater than those produced by conventional deadline-aware protocols.

To further verify if the proposed protocol always produces more data importance, we analyze the delivered flows at the aggregating host. Figure \ref{fig:ratio} shows the ratio of flows that successfully meet their deadline in the heavy-load network. The delivery ratio of important flows of the proposed protocol significantly outperforms others. This proves that our importance-aware protocol can improve the delivery ratio of most important flows.


\subsection{Results of NTCIR dataset}

\begin{figure}
  \centering
  \includegraphics[width=95mm,trim=35 0 20 0, clip]{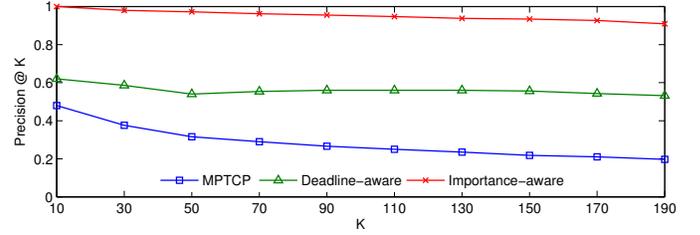}\\
  \caption{Precision at different K}\label{fig:topk_t}
\end{figure}

In this scenario, we first use a common retrieval model, i.e. the vector space model, to calculate the similarity of each data unit to each query. Then, we can obtain the top-$K$ relevant rank lists of each query. In each rank list of a query, $K$ data units, which are most similar to the query, are recorded as the ground truth. Hence, we use the data units received by the aggregating host before deadline and the rank list to calculate the precision at $K$ of the query. Figure \ref{fig:topk_t} gives the results of the precision at $K$ when the deadlines are tight. Not surprisingly, the proposed protocol significantly outperforms MPTCP and D$^{3}$. MPTCP is unable to satisfy the requirements of deadline-aware services because of the RTO problem. D$^{3}$ can not achieve high precision because it does not consider flow importance. The proposed protocol, as mentioned before, always delivers most important flows first, and hence provides outstanding precision at $K$ all the time.

\section{Conclusion}
In this work, we investigate the requirements to provide high quality deadline-aware online services in datacenters to users. Based on the observations from the literatures\cite{lagun2011viewser, buscher2010good, cutrell2007you, lorigo2008eye} and the observation of flow contents, we found that the fulfillment of deadline constraint and high bandwidth utilization do not guarantee high service quality. Thus, we exploit the application-level content information, i.e., data importance, to better fulfill the requirements of deadline-aware services for providing better user experience, and hence producing more revenue for the service providers.

To figure out the optimal solution, we first model the datacenter network as a flow network for analyzing the application importance maximization problem with the constraints of network link capacity, flow conservation, and flow deadline. Although the property of static multi-disjoin routing paths, which exists in low-cost server-centric datacenter infrastructure designs i.e., BCube\cite{bcube} and MDCube\cite{mdcube}, significantly reduces the complexity of the problem, the optimal solution is still impractical for realistic datacenter networks.

Therefore, based on the ideas of the optimal solution, we further propose a novel distributed rate-based importance-aware delivery control protocol. The results shows that our proposed protocol provides significant benefits over even optimized versions of existing protocols in terms of both applicaion-level throughput and importance. The results of real-world dataset also demonstrate that the proposed protocol can provide better precision at $K$ all the time. Since the growing demand for deadline-aware online services, we believe exploiting data importance to provide high-rank results for users is important toward high-quality deadline-aware services beyond the state-of-art solutions.

\bibliographystyle{IEEEtran}


\begin{thebibliography}{10}
\providecommand{\url}[1]{#1}
\csname url@samestyle\endcsname
\providecommand{\newblock}{\relax}
\providecommand{\bibinfo}[2]{#2}
\providecommand{\BIBentrySTDinterwordspacing}{\spaceskip=0pt\relax}
\providecommand{\BIBentryALTinterwordstretchfactor}{4}
\providecommand{\BIBentryALTinterwordspacing}{\spaceskip=\fontdimen2\font plus
\BIBentryALTinterwordstretchfactor\fontdimen3\font minus
  \fontdimen4\font\relax}
\providecommand{\BIBforeignlanguage}[2]{{%
\expandafter\ifx\csname l@#1\endcsname\relax
\typeout{** WARNING: IEEEtran.bst: No hyphenation pattern has been}%
\typeout{** loaded for the language `#1'. Using the pattern for}%
\typeout{** the default language instead.}%
\else
\language=\csname l@#1\endcsname
\fi
#2}}
\providecommand{\BIBdecl}{\relax}
\BIBdecl

\bibitem{lagun2011viewser}
D.~Lagun and E.~Agichtein, ``Viewser: Enabling large-scale remote user studies
  of web search examination and interaction,'' in \emph{ACM SIGIR}, 2011.

\bibitem{buscher2010good}
G.~Buscher, S.~Dumais, and E.~Cutrell, ``The good, the bad, and the random: an
  eye-tracking study of ad quality in web search,'' in \emph{Proceeding of the
  33rd international ACM SIGIR conference on Research and development in
  information retrieval}.\hskip 1em plus 0.5em minus 0.4em\relax ACM, 2010, pp.
  42--49.

\bibitem{cutrell2007you}
E.~Cutrell and Z.~Guan, ``What are you looking for?: an eye-tracking study of
  information usage in web search,'' in \emph{Proceedings of the SIGCHI
  conference on Human factors in computing systems}.\hskip 1em plus 0.5em minus
  0.4em\relax ACM, 2007.

\bibitem{lorigo2008eye}
L.~Lorigo, M.~Haridasan, H.~Brynjarsd{\'o}ttir, L.~Xia, T.~Joachims, G.~Gay,
  L.~Granka, F.~Pellacini, and B.~Pan, ``Eye tracking and online search:
  Lessons learned and challenges ahead,'' \emph{Journal of the American Society
  for Information Science and Technology}, 2008.

\bibitem{ntcir}
NTCIR, ``http://research.nii.ac.jp/ntcir/index-en.html.''

\bibitem{gfs}
S.~Ghemawat, H.~Gobioff, and S.-T. Leung, ``The google file system,''
  \emph{SIGOPS Oper. Syst. Rev.}, 2003.

\bibitem{deadline}
C.~Wilson, H.~Ballani, T.~Karagiannis, and A.~Rowtron, ``Better never than
  late: meeting deadlines in datacenter networks,'' in \emph{Proceedings of the
  ACM SIGCOMM 2011 conference}, ser. SIGCOMM '11.\hskip 1em plus 0.5em minus
  0.4em\relax New York, NY, USA: ACM, 2011, pp. 50--61.

\bibitem{d2tcp}
T.~N.~V. Balajee~Vamanan, Jahangir~Hasan, ``Deadline-aware datacenter tcp
  (d2tcp),'' ser. SIGCOMM'12.\hskip 1em plus 0.5em minus 0.4em\relax Purdue,
  West Lafayette, USA: ACM, 2012.

\bibitem{bcube}
C.~Guo, G.~Lu, D.~Li, H.~Wu, X.~Zhang, Y.~Shi, C.~Tian, Y.~Zhang, and S.~Lu,
  ``Bcube: a high performance, server-centric network architecture for modular
  data centers,'' \emph{SIGCOMM Comput. Commun. Rev.}, 2009.

\bibitem{mdcube}
H.~Wu, G.~Lu, D.~Li, C.~Guo, and Y.~Zhang, ``Mdcube: a high performance network
  structure for modular data center interconnection,'' in \emph{Proceedings of
  the 5th international conference on Emerging networking experiments and
  technologies}.\hskip 1em plus 0.5em minus 0.4em\relax ACM, 2009, pp. 25--36.

\bibitem{mptcp}
C.~Raiciu, C.~Pluntke, S.~Barre, A.~Greenhalgh, D.~Wischik, and M.~Handley,
  ``Data center networking with multipath tcp,'' in \emph{Proceedings of the
  9th ACM SIGCOMM Workshop on Hot Topics in Networks}, ser. Hotnets-IX.\hskip
  1em plus 0.5em minus 0.4em\relax New York, NY, USA: ACM, 2010, pp.
  10:1--10:6.

\bibitem{approximationunsplittable}
A.~Chakrabarti, C.~Chekuri, A.~Gupta, and A.~Kumar, ``Approximation algorithms
  for the unsplittable flow problem,'' in \emph{Proceedings of the 5th
  International Workshop on Approximation Algorithms for Combinatorial
  Optimization}, ser. APPROX '02.\hskip 1em plus 0.5em minus 0.4em\relax
  London, UK, UK: Springer-Verlag, 2002, pp. 51--66.

\bibitem{reducibility}
R.~M. Karp, ``{Reducibility among combinatorial problems" in complexity of
  computer computations},'' 1972.

\bibitem{ns2}
``The network simulator ns-2, \\ http://www.isi.edu/nsnam/ns/.''

\bibitem{mptcpns2}
``Multiple-path tcp project, \\ http://code.google.com/p/multipath-tcp/.''

\bibitem{incast2}
Y.~Chen, R.~Griffith, J.~Liu, R.~H. Katz, and A.~D. Joseph, ``Understanding tcp
  incast throughput collapse in datacenter networks,'' in \emph{Proceedings of
  the 1st ACM workshop on Research on enterprise networking}, ser. WREN
  '09.\hskip 1em plus 0.5em minus 0.4em\relax New York, NY, USA: ACM, 2009.

\bibitem{websearchforaplanet}
L.~Barroso, J.~Dean, and U.~Holzle, ``Web search for a planet: The google
  cluster architecture,'' \emph{Micro, IEEE}, march-april 2003.

\bibitem{mapreduce}
J.~Dean and S.~Ghemawat, ``Mapreduce: simplified data processing on large
  clusters,'' \emph{Commun. ACM}, vol.~51, no.~1, pp. 107--113, Jan. 2008.

\bibitem{dctcp}
M.~Alizadeh, A.~Greenberg, D.~A. Maltz, J.~Padhye, P.~Patel, B.~Prabhakar,
  S.~Sengupta, and M.~Sridharan, ``Data center tcp (dctcp),'' in
  \emph{Proceedings of the ACM SIGCOMM 2010 conference}, ser. SIGCOMM
  '10.\hskip 1em plus 0.5em minus 0.4em\relax New York, NY, USA: ACM, 2010, pp.
  63--74.

\bibitem{ictcp}
H.~Wu, Z.~Feng, C.~Guo, and Y.~Zhang, ``Ictcp: Incast congestion control for
  tcp in data center networks,'' in \emph{Proceedings of the 6th International
  COnference}, ser. Co-NEXT '10.\hskip 1em plus 0.5em minus 0.4em\relax New
  York, NY, USA: ACM, 2010.

\bibitem{hull}
M.~Alizadeh, A.~Kabbani, T.~Edsall, B.~Prabhakar, A.~Vahdat, and M.~Yasuda,
  ``Less is more: trading a little bandwidth for ultra-low latency in the data
  center,'' in \emph{Proceedings of the 9th USENIX conference on Networked
  Systems Design and Implementation}, ser. NSDI'12.\hskip 1em plus 0.5em minus
  0.4em\relax Berkeley, CA, USA: USENIX Association, 2012, pp. 19--19.

\end{thebibliography}
%




\end{document}